\documentclass[12pt,preprint]{aastex}
\usepackage[]{natbib}

\received{}
\accepted{}
\journalid{}{}
\articleid{}{}
\begin{document}
\def\ro{{\it ROSAT\/}}
\def\xmm{{\it XMM-Newton}}
\def\chandra{{\it Chandra}}
\def\hess{HESS~J1303$-$631}
\def\hegra{TeV~J2032+4130}

\title{\chandra\ Observation of the Unidentified TeV Gamma-Ray
Source \hess\ in the Galactic Plane}

\author {R. Mukherjee\altaffilmark{1} \& J. P. Halpern\altaffilmark{2}}
\altaffiltext{1}
{Department of Physics \& Astronomy, Barnard College, Columbia University,
New York, NY 10027}
\altaffiltext{2}
{Department of Astronomy, Columbia University, New York, NY 10027}

\begin{abstract}
The imaging atmospheric Cherenkov array H.E.S.S. recently
discovered an extended source in the 0.4$-$10~TeV energy range,
\hess.  We obtained a 5~ks observation with the ACIS-I array on the
\chandra\ X-ray observatory that
does not reveal an obvious compact or diffuse X-ray counterpart.
Archival \ro\ images are also blank in this region.  Although
there are several radio pulsars within the field of \hess, none
is detected in X-rays to a flux limit of $<5 \times 10^{-14}$
ergs~cm$^{-2}$ s$^{-1}$, and none is a likely counterpart
on energetic grounds.  Over the entire $17^{\prime} \times 17^{\prime}$
ACIS-I field, we place an upper limit of $<5.4 \times 10^{-12}$
ergs cm$^{-2}$ s$^{-1}$ on the excess diffuse flux in the $2-10$~keV band.
One hard point-source with flux
$\approx 4 \times 10^{-14}$ ergs~cm$^{-2}$ s$^{-1}$
lies within $0.\!^{\prime}5$ of the centroid of the TeV emission.
These exploratory observations suggest that deeper
pointings with \chandra\ and \xmm\ are needed before we can learn more
about the nature of \hess.  Its similarity to the unidentified
source \hegra\ indicates the probable existence of a new class of
high-energy source in the Galactic plane that originates from young,
massive stars or their supernova remnants.

\end{abstract}

\newpage

\section{Introduction}

The new imaging atmospheric Cherenkov telescope (IACT) H.E.S.S.
(High Energy Stereoscopic System) is located in Namibia and
is sensitive in the hundreds of GeV regime \citep{hin04}.
The complete H.E.S.S. array of four telescopes has been operational
since 2003 December and has reported observations of several astrophysical
sources \citep[e.g.,][]{aha04,bei04a}.  Recent results
include the discovery of a new,
unidentified TeV $\gamma$-ray source \hess\ \citep{bei04a,aha05a}.
This is only the second unidentified source
at TeV energies in the Galactic plane,
at $(l,b)$ = ($304.\!^{\circ}24$, $-0.\!^{\circ}36$).
\hess\ was discovered in the pointings at the
Be/X-ray pulsar PSR B1259$-$63 \citep{bei04b} for 39 hr
of total exposure from 2004 February to June. The new source is
located $0.\!^{\circ}6$ north of the pulsar, at R.A.=
$13^{\rm h}03^{\rm m}00.\!^{\rm s}4 \pm 04.\!^{\rm s}4$, decl.=
$-63^{\circ}11^{\prime}55^{\prime\prime} \pm 31^{\prime\prime}$
(J2000.0), and was observed to be brighter than the pulsar, about 0.17
crab, or $1.2 \times 10^{-11}$ ergs~cm$^{-2}$~s$^{-1}$ above 380~GeV. 
The statistical significance of detection was $21 \sigma$.
There is good evidence that the TeV emission is extended, 
with an intrinsic Gaussian $\sigma=0.\!^{\circ}16 \pm 0.\!^{\circ}02$.
There is no evidence of flux variability
among the H.E.S.S. observations.
These properties are similar to those of TeV J2032+4130, the first such
unidentified source that was discovered by the HEGRA CT system \citep{aha02}.
Also in the Galactic plane at $(l,b)$ = ($80.\!^{\circ}25$, $+1.\!^{\circ}07$), 
\hegra\ lies in the direction of the Cygnus OB2 
stellar association, and its nature remains uncertain
despite extensive multiwavelength efforts to
find a counterpart \citep{muk03,but03}. 

In this paper we report on initial efforts to identify \hess.
Multiwavelength searches of individual EGRET source fields were
often successful in finding a likely
identification \citep[see][for a review]{muk04}.
A difficulty faced by the EGRET searches,
especially in the Galactic plane,
is the relatively large positional uncertainty,
$\approx 0.\!^{\circ}5-1^{\circ}$.
IACTs like H.E.S.S. or HEGRA have the advantage of much
better angular resolution resulting in smaller error boxes for
$\gamma$-ray source positions in comparison to EGRET,
even when the extent of \hess\ is included.

\section{High-Energy and Radio Observations}

\subsection{\ro }

\begin{figure}[t]
  \begin{center}
    \includegraphics[height=37pc]{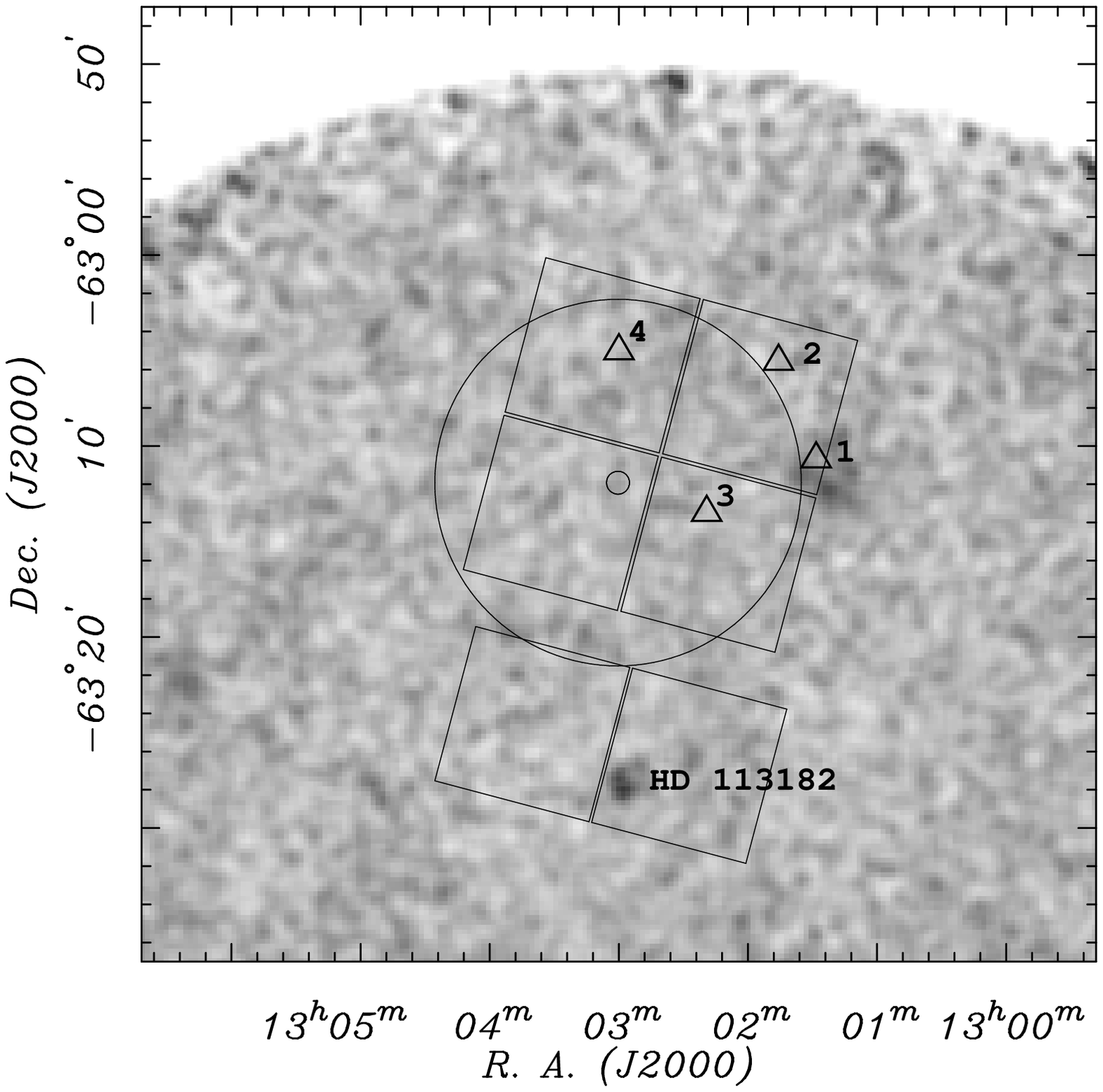}
  \end{center}
  \caption{{\sl ROSAT\/} PSPC X-ray image of the field of \hess.
The small circle 
corresponds to the $1\sigma$ uncertainty of the centroid of the 
H.E.S.S. source, which includes both statistical
errors and pointing (systematic) errors. 
The large circle is the estimated Gaussian $1 \sigma$ extent of the TeV emission
\citep{aha05a}. The squares are the fields of view of the CCDs in the subsequent
\chandra\ observation (see Figure~\ref{chandra_fig}).
The numbered triangles are the locations of radio pulsars from
the Parkes Multibeam Survey, whose properties are listed in Table~\ref{pulsar_tab}.
}
\label{rosat_fig}
\end{figure}

Prior to the \chandra\ observation reported here, the error box of
\hess\ was covered only by a {\it ROSAT\/} observation,
which does not show any bright sources in the region of
interest. No pointings by {\it ASCA\/} or {\it BeppoSAX\/} were made in
this region. Since this source is in the Galactic plane (similar to \hegra),
it is likely that its X-ray counterpart is highly absorbed.
The total \ion{H}{1} column density in this direction is
$N_{\rm H} = 2 \times 10^{22}$ cm$^{-2}$, which, not even
including molecular hydrogen, has optical depth 
$> 1$ at $E < 3$~keV, and should be associated with
visusal extinction  $A_V \approx 11$~mag \citep{pre95}.
Figure~\ref{rosat_fig} shows the X-ray image taken with the {\it ROSAT\/}
Position Sensitive Proportional Counter (PSPC)
in the energy range 0.2--2.0 keV, covering the field of
\hess. This image was created by co-adding
exposure-corrected skymaps of 49 ks of data.
The original target of the {\it ROSAT\/} observations was PSR B1259$-$63. 
The small circle corresponds to the $1\sigma$
uncertainty of the centroid of \hess, and includes both statistical errors and
pointing (systematic) errors. The systematic pointing uncertainty of the
H.E.S.S. telescopes is estimated to be 
$\sim 20^{\prime\prime}$ \citep{aha05a}. The statistical and systematic errors
were added in quadrature. 
The large circle around the TeV position indicates the 
Gaussian $\sigma = 0.\!^{\circ}16$
extent of the TeV source \citep{aha05a}.  The only significant
X-ray source in Figure~\ref{rosat_fig} is the K0 star HD~113182 at 
R.A. = $13^{\rm h}02^{\rm m}59.\!^{\rm s}3$,
decl. = $-63^{\circ}27^{\prime}44^{\prime\prime}$ (J2000.0),
which is inconsistent with the H.E.S.S. location.
The numbered triangles in the region of the H.E.S.S. source are
the locations of radio pulsars from the Parkes Multibeam Survey.
Their properties are given in Table~\ref{pulsar_tab}.
While there are several radio pulsars near the
H.E.S.S. position, none is detected in the {\it ROSAT\/} image.
A weak, out-of-focus {\it ROSAT\/} source near the location of
PSR J1301$-$6310 ($d \approx 2.1$~kpc)
is detected only in soft X-rays, indicating
that it is probably from a foreground star, and 
not attributable to the pulsar.  Within the Gaussian $\sigma$
extent of the H.E.S.S. source, there are no other candidate
radio sources in published surveys \citep{aha05a}.

\subsection{\chandra }

\begin{figure}
  \begin{center}
\includegraphics[height=35pc]{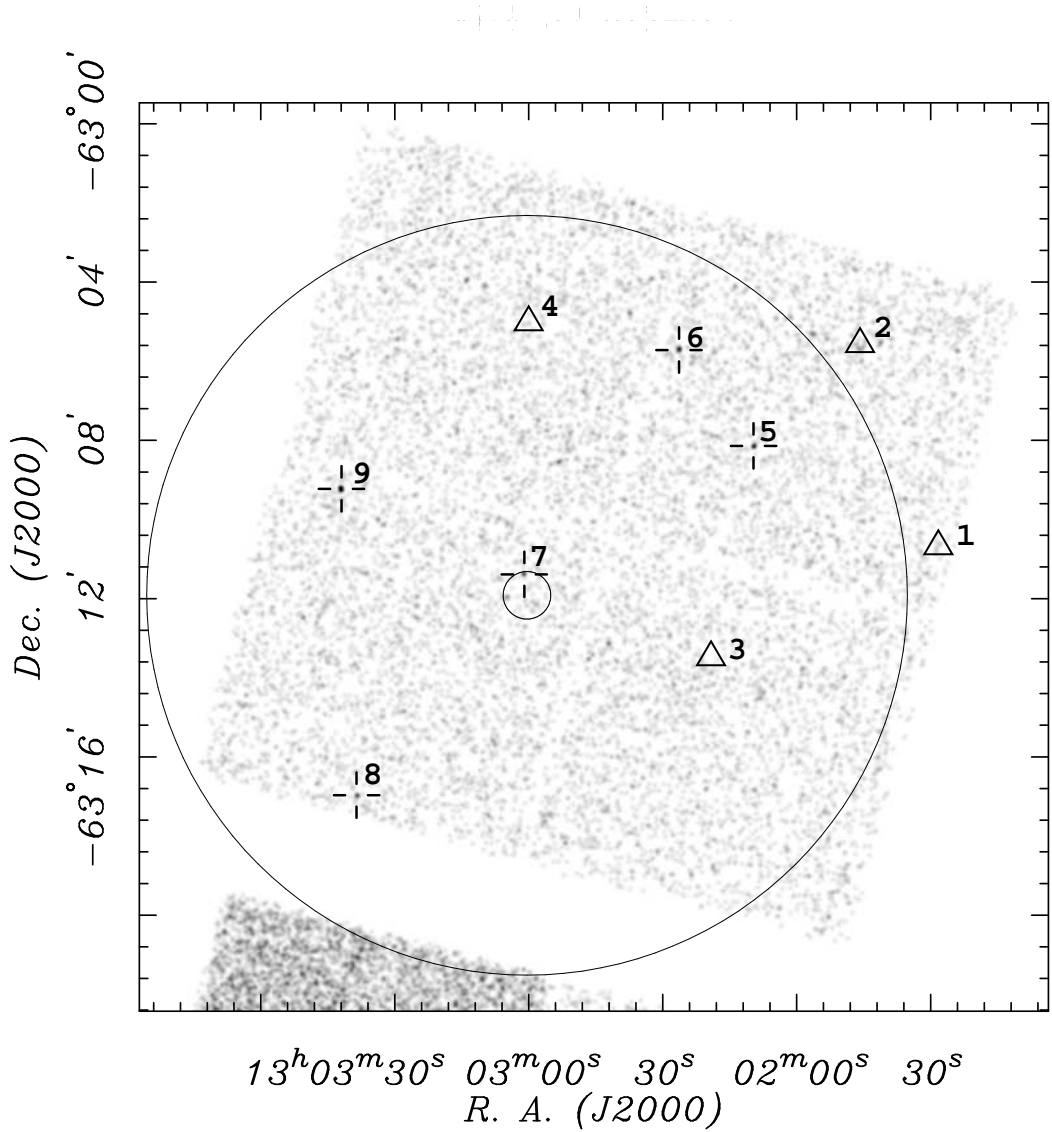}
  \end{center}
\vspace{-0.5cm}
  \caption{\chandra\ ACIS-I image of the field of \hess.
The numbered triangles mark the locations of radio pulsars from
the Parkes Multibeam Survey, whose properties are listed in
Table~\ref{pulsar_tab}. The small circle 
corresponds to the $1\sigma$
uncertainty of the centroid of \hess, and includes both statistical errors and
pointing (systematic) errors. The large circle is the estimated
Gaussian $1 \sigma$ extent of the TeV emission \citep{aha05a}.
The plus signs mark the significant \chandra\ point sources, whose
properties are listed in Table~\ref{chandra_tab}. Source 7 is
$32^{\prime\prime}$ from the nominal centroid of the TeV emission. }
\label{chandra_fig}
\end{figure}

We observed the location of \hess\ on 2004 September 25 
with the Advanced CCD Imaging Spectrometer \citep[ACIS,][]{bur97} onboard the
\chandra\ X-ray Observatory \citep*{wei96,wei02} using
Director's discretionary time.  A preliminary estimate of the source centroid
was positioned at the default location on the front-illuminated ACIS-I array.
The standard timed readout with a frame time
of 3.2~s was used, and the data were
collected in VFAINT mode.  A total of 4707~s of on-time was
accumulated, while the effective exposure live-time was 4648~s.
After processing the event file through
the VFAINT mode filter, a total of 4188 diffuse events in the $2-10$~keV band
were found on the four ACIS-I CCDs in addition to five point sources with
seven or more photons each.
Figure~\ref{chandra_fig} shows the \chandra\ image with the five point
sources marked. Their positions and count rates are listed in Table~\ref{chandra_tab},
together with possible stellar optical counterparts.
None of the radio pulsars listed in Table~\ref{pulsar_tab} was detected in the
{\it Chandra\/} observation, to a limit of $ \approx 5 \times 10^{-14}$
ergs~cm$^{-2}$ s$^{-1}$ in the $2-10$~keV band.

The one hard X-ray source (\#7) without a possible optical (or 2MASS) counterpart
lies only $32^{\prime\prime}$ from the centroid of the TeV source, which
is consistent with the combined $20^{\prime\prime}$ systematic and 
$31^{\prime\prime}$ statistical uncertainties of the latter.
Assuming a power-law spectrum of photon index $\Gamma = 2.0$ and column
density $N_{\rm H} = 2 \times 10^{22}$ cm$^{-2}$ (the total \ion{H}{1} column
density in this direction), source 7 has a $2-10$~keV flux of 
$\approx 4 \times 10^{-14}$ ergs cm$^{-2}$ s$^{-1}$.  It could be a
background AGN, or a candidate for identification with \hess.

Any fainter point sources 
than the ones listed in Table~\ref{chandra_tab} are of marginal
significance, and are included in the following analysis of the diffuse flux.
Although no extended source is apparent in the {\it Chandra\/} image, we
use the standard quiescent background
rates to estimate that 3239 of the diffuse events in the $2-10$~keV
band are particle plus cosmic
X-ray background, leaving 949 diffuse photons over the
$17^{\prime} \times 17^{\prime}$ ACIS-I field.  These 949 photons may be
Galactic diffuse background or diffuse flux originating from \hess.
Conservatively, this 0.21 photons s$^{-1}$ of diffuse flux can be considered an
upper limit to the flux of \hess.  Assuming a power-law
spectrum of photon index $\Gamma = 2.0$ and column density
$N_{\rm H} = 2 \times 10^{22}$ cm$^{-2}$, this corresponds to a $2-10$~keV
observed flux of $<5.4 \times 10^{-12}$ ergs cm$^{-2}$ s$^{-1}$, or an
unabsorbed $2-10$~keV flux of $<6.4 \times 10^{-12}$ ergs cm$^{-2}$ s$^{-1}$.

If we perform the same analysis separately in the $2-5$~keV band,
where the instrumental background is lower and the subtraction less
subject to systematic effects, we find an observed diffuse
flux of  $<2.2\times 10^{-12}$ ergs cm$^{-2}$ s$^{-1}$, or an
unabsorbed $2-5$~keV flux of $<3.0 \times 10^{-12}$ ergs cm$^{-2}$ s$^{-1}$.
These limits would be reduced by a factor of several if the diffuse source
is considerably smaller than the $17^{\prime} \times 17^{\prime}$ ACIS-I field.
The fluxes limits are subject to additional uncertainty from the unknown shape 
of the X-ray spectrum, and possible column density of intervening
molecular material would increase the limits.

\subsection{EGRET}

\hess\ was not detected by EGRET above 100 MeV. The source is located in
the Galactic plane, which has intense diffuse gamma-ray emission.
For the Galactic plane sources, the third EGRET (3EG) Catalog \citep{har99}
lists point sources that were detected at least $5\sigma$
above the diffuse model. We have estimated the upper limit of $\gamma$-ray
emission to be $10.4\times 10^{-8}$ photons
cm$^{-2}$ s$^{-1}$ above 100 MeV.

\section{Discussion}

The short \chandra\ observation presented here does not reveal much 
about the nature of \hess. Those point sources detected 
primarily in soft X-rays ($< 2$~keV) are likely to be coronal,
with late-type stellar counterparts identified in Table~\ref{chandra_tab},
and unrelated to \hess.  The one hard X-ray source
without an optical counterpart (\#7 in Table~\ref{chandra_tab})
could be a background AGN.  The position of source 7 is consistent
with the well-determined ($0.\!^{\prime}5$)
location of the centroid of the TeV emission.
However, the extended and steady nature of \hess\
disfavors a blazar origin.  The absence of a brighter X-ray counterpart
argues against a microquasar origin, but perhaps allows an
isolated compact object.

No diffuse source is apparent
in the \chandra\ image, thus, there is no
further evidence concerning the extended TeV emission,
and whether it has an electronic or hadronic origin.
Based on their spin-down fluxes $\dot E/4\pi d^2$, none of the known pulsars
in the \hess\ field (Table~\ref{pulsar_tab}) is a good candidate for powering
detectable TeV emission from a pulsar wind nebula (PWN).
The highest spin-down flux is from PSR J1301$-$6305, 
which has $\dot E/4\pi d^2 = 5.6 \times 10^{-11}$
ergs~cm$^{-2}$~s$^{-1}$, and would imply an unreasonably high efficiency
of $\approx 38\%$ for production of $0.3-10$ TeV flux,
which is $2.1 \times 10^{-11}$
ergs~cm$^{-2}$~s$^{-1}$ according to the extrapolated spectrum \citep{aha05a}.
In comparison, \chandra\ sets an
upper limit on the 2$-$10 keV flux of each of these pulsars of
$\approx 5 \times 10^{-14}$ ergs~cm$^{-2}$~s$^{-1}$, which is consistent
with the normal range of $10^{-4} \leq L_{\rm x}/\dot E \leq 10^{-2}$ for
pulsars \citep{pos02}.  
Furthermore, there is an empirical threshold $\dot E$ of
$\approx 4 \times 10^{36}$ ergs~s$^{-1}$ below which pulsars do not
power bright PWN either in X-rays or in radio \citep{got04}.
None of the pulsars in Table~\ref{pulsar_tab}, including PSR J1301$-$6305,
exceeds this threshold.

It is instructive to compare this \chandra\
observation of \hess\ to a similar 5~ks \chandra\ observation of the HEGRA
source \hegra,
the first unidentified TeV source. 
\hegra\ is located near the Cygnus OB2 region, in which several of the stars
are among the most active stellar
X-ray sources in the Galaxy.  In contrast
to \hess, the \hegra\ field contains more point X-ray
sources, spectroscopic optical identifications of which indicated
that they are mostly a mix of O-type stars in the Cyg OB2
association, at $d \approx 1.7$~kpc,
and foreground late-type stars \citep{muk03}.
The only unusual X-ray source in the field of \hegra\ is a
highly variable, hard and absorbed X-ray
source, the brightest in the \chandra\ image, located
$7^\prime$ from the centroid of \hegra.
It is not clear if this unidentified X-ray
source is associated with \hegra. If so, \citet{muk03}
argued that it would be located considerably beyond Cyg OB2,
and still be within the young Galactic disk.

It has been argued that colliding
stellar winds in the Cyg OB2 association could be responsible for
the production of TeV gamma-rays in \hegra\ \citep{ben01}. \citet*{aha02}
hypothesized two possible origins for extended TeV emission
that may be displaced from its originating source of energy. One is 
$\pi^0$ decay resulting 
from hadrons accelerated in shocked OB star winds and interacting
with a local, dense cloud. The other is TeV
emission in a jet-driven termination shock due to Inverse Compton scattering,
either from an as-yet undetected microquasar, or from Cyg X-3.
\citet*{tor04} theorized that groups of stars in OB associations 
located close to cosmic ray acceleration sites may be sources
of TeV emission, and cited
\hegra\ as a candidate for such a source. In such a scenario,
no MeV-GeV emission is expected from the source at EGRET or
{\it GLAST\/} energies.

The HEGRA source 
\hegra\ also showed some evidence for spatial extent, with a Gaussian $1\sigma$
radius of $5.\!^{\prime}6 \pm 1.\!^{\prime}7$ \citep{aha02}. As in the case of
\hess, the HEGRA source was found to be steady in observations
from 1999 to 2001, unlike the flaring blazars detected by EGRET and the
IACTs.  However,
unlike \hegra, \hess\ is not located near a dense OB stellar association 
or a star formation region.
Although \citet{aha05a} noted that \hess\ is superposed on the
Centaurus OB1 association, this is a rather sparse group
of only 20 stars, mostly of type B, spread over $\approx 6^{\circ}$ or 260 pc 
at the adopted distance of 2.5 kpc \citep{hum78} (2.3~kpc
according to \citealt{kal94}).
The nearest member is $0.\!^{\circ}8$ from \hess.  Therefore,
we hesitate to posit a physical origin of the small TeV
source from the group as a whole, or even from any existing member.
Also, the average visual extinction to stars in Cen OB1 is only
$A_V \approx 2.0$~mag \citep{kal94}, which is not large enough to account 
for the absence of an X-ray counterpart at this distance.
However, it is possible that a deceased member of the association,
whose compact remnant is difficult to detect, is responsible for \hess.

\section{Conclusions}

That both \hegra\ and \hess\ are indeed Galactic now seems secure.
It is likely that several more such sources will be detected in future sky
surveys to be conducted by H.E.S.S. and VERITAS.  Such TeV sources may not
be associated with the numerous unidentified EGRET sources in the 3EG
catalog, and could very well comprise a new class by themselves.
Given the apparent latitude range of $\sim 1^{\circ}$,
their scale height must be $<200$~pc if located
at distances $<10$~kpc.  Thus, they are not very old, and
probably originate from massive stars or their remnants.

After this paper was submitted, the discovery by H.E.S.S. of eight
additional sources in a survey of the Galactic plane was reported \citep{aha05b}.
Their mean Galactic latitude of $-0.\!^{\circ}25$ and rms dispersion
in latitude of $0.\!^{\circ}25$
further supports an association with young, massive stars.
All of these new sources appear to be extended, and some are plausibly associated
with known supernova remnants, which suggests that the remaining unidentified
sources might also be supernova remnants that have not yet been detected
or recognized at longer wavelengths because of their considerable distances
and/or weak synchrotron emission.
The search for and evaluation of candidate counterparts
will be very important in resolving the nature of these TeV sources.

\acknowledgements

We thank Harvey Tananbaum for granting the \chandra\ Director's discretionary
time for this observation. This work was supported in part by the National
Science Foundation.

\begin{deluxetable}{lcclccrcc}
\rotate
\tablenum{1}
\tablecolumns{6}
\tablewidth{0pc}
\tablecaption{Radio Pulsars in the Field of \hess}
\tablehead
{
Ident. & Name & Position &  \omit\hfil $P$ \hfil &
$\log \tau$ & $\log \dot E$ & \omit\hfil $d$ \hfil & $\log (\dot E/4\pi d^2)$ & Ref \\
& & R.A. \hskip 0.5in Decl. & \omit\hfil (s) \hfil & \omit\hfil (yr) \hfil & (ergs s$^{-1}$) &
(kpc) & (ergs cm$^{-2}$ s$^{-1}$)
}
\startdata
1 & PSR J1301$-$6310 & 13 01 28.3  --63 10 40 & 0.664 & 5.27 & 33.88 & 2.1  & --10.83 & 1 \\
2 & PSR J1301$-$6305 & 13 01 45.8  --63 05 34 & 0.185 & 4.04 & 36.23 & 15.8 & --10.25 & 2  \\
3 & PSR J1302$-$6313 & 13 02 19.2  --63 13 29 & 0.968 & 6.38 & 32.44 & 28.1 & --14.53 & 1 \\
4 & PSR J1303$-$6305 & 13 03 00.0  --63 05 01 & 2.307 & 7.22 & 30.85 & 13.6 & --15.50 & 2 \\
\enddata
\tablecomments{Units of right ascension are hours, minutes, and seconds.
Units of declination are degrees, arcminutes, and arcseconds.}
\tablerefs{(1) \citealt{kra03}; (2) \citealt{man01}.}
\label{pulsar_tab}
\end{deluxetable}

\clearpage
\begin{deluxetable}{lccccccc}
\rotate
\tablenum{2}
\tablecolumns{8}
\tablewidth{0pc}
\tablecaption{\chandra\ Sources in the Field of \hess\ }
\tablehead
{
\chandra\ &  X-ray Position & Counts
& Counts & Nearest USNO B1.0 Star & $B$ & $R$
& Dist\\
Ident. & R.A.\hskip 0.5in Decl. & ($< 2$ keV) & ($> 2$ keV) & R.A.\hskip 0.5in
Decl. & (mag) & (mag) & ($^{\prime\prime}$)
}
\startdata
5& 13 02 09.67 --63 08 08.4  &  9 &0 & 13 02 09.53 --63 08 09.1 &  18.40 & 16.02 &  1.2\\
6& 13 02 26.35 --63 05 43.0  &  9 &0 & 13 02 26.42 --63 05 42.4 &  18.59 & 16.10 &  0.8\\
7& 13 03 00.98 --63 11 23.2  &  0 &7 &     ... \hskip 0.5in ...    &  ...  &  ...  &  ...\\
8& 13 03 38.59 --63 16 58.5  &  7 &0 & 13 03 38.71 --63 16 59.2 &  15.01 & 14.32 &  1.1\\
9& 13 03 41.93 --63 09 13.5  & 13 &5 & 13 03 41.92 --63 09 12.9 &  19.68 & 16.52 &  0.6\\
 &                           &    &  & 13 03 42.11 --63 09 13.8 &  16.46 & 15.78 &  1.2\\
\enddata
\tablecomments{Units of right ascension are hours, minutes, and seconds.
Units of declination are degrees, arcminutes, and arcseconds.}
\label{chandra_tab}
\end{deluxetable}

\end{document}